\documentclass{iaus}
\usepackage{graphicx}

\title[Dynamo Models of the Solar Cycle]
{Modeling the Solar Cycle: What the Future Holds}

\author[Dibyendu Nandy]
{Dibyendu Nandy}

\affiliation{Indian Institute of Science Education and Research, Kolkata, \\ Mohanpur 741252, West Bengal, India,  \\email: {\tt dnandi@iiserkol.ac.in}}

\pubyear{2012}
\volume{286}  
\pagerange{1--10}
\jname{Comparative Magnetic Minima: Characterizing quiet times in the Sun and stars}
\editors{C.H. Mandrini \& D. Webb}

\begin{document}

\maketitle

\begin{abstract}

Stellar magnetic fields are produced by a magnetohydrodynamic dynamo mechanism working in their interior -- which relies on the interaction between plasma flows and magnetic fields. The Sun, being a well-observed star, offers an unique opportunity to test theoretical ideas and models of stellar magnetic field generation. Solar magnetic fields produce sunspots, whose number increases and decreases with a 11 year periodicity -- giving rise to what is known as the solar cycle. Dynamo models of the solar cycle seek to understand its origin, variation and evolution with time. In this review, I summarize observations of the solar cycle and describe theoretical ideas and dynamo modeling efforts to address its origin. I end with a discussion on the future of solar cycle modeling -- emphasizing the importance of a close synergy between observational data assimilation, kinematic dynamo models and full magnetohydrodynamic models of the solar interior.


\end{abstract}

\firstsection

\section{Introduction to the Solar Cycle}

\begin{figure}[h!]
\begin{center}
\hspace{-1.0cm}
\includegraphics[width=7.2cm]{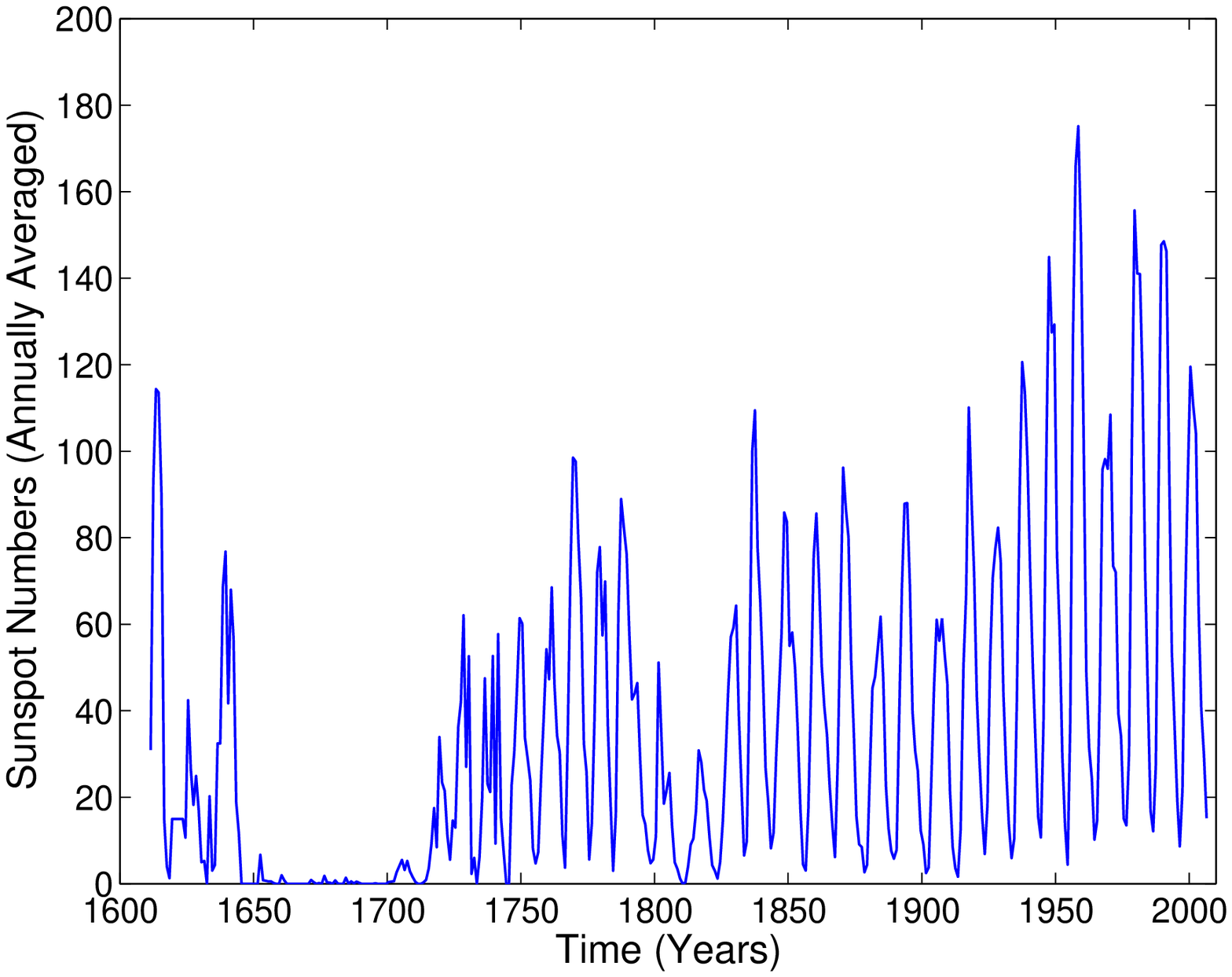}
\end{center}
\begin{center}
\includegraphics[width=10.5cm]{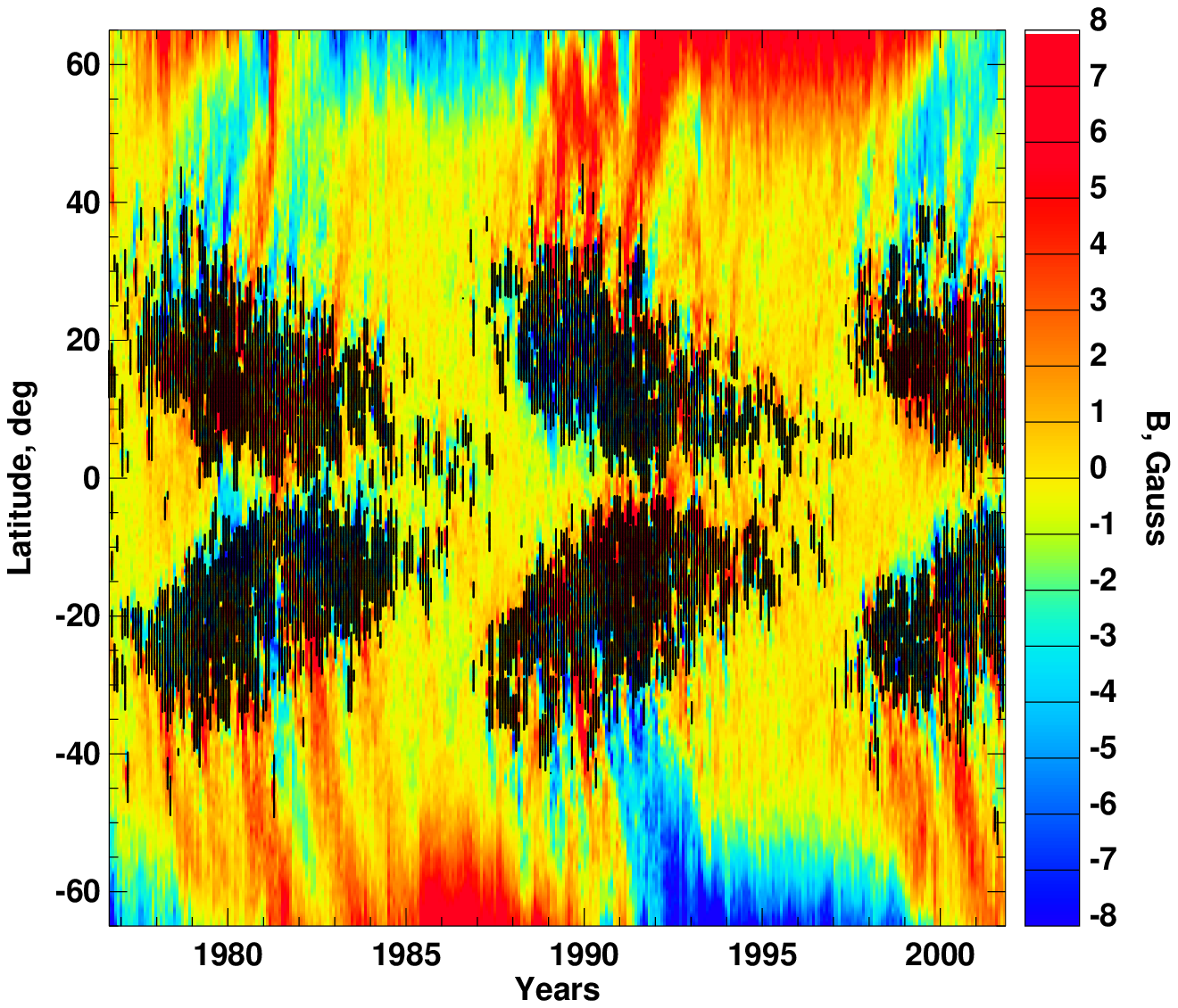}
\caption{Top: The variation in the number of sunspots observed on the Sun with time over the last fours centuries. The variation is for the most part, cyclic, ranging typically from 9--14 year cycles and with an overall average period of 11 years. The amplitude of the cycle is observed to fluctuate significantly from one cycle to another. A notable period is the Maunder minimum between 1645--1715 AD, when the sunspot cycle pretty much stopped. Solar activity reconstruction based on cosmogenic isotopes indicate that there have been many such episodes of reduced or no activity. Bottom: A butterfly diagram (latitude-time plot) of solar surface magnetic field showing the evolution of latitudes where sunspots emerge (black regions) and the surface radial field (outside of sunspots; scale on right-hand y-axis) with the progress of time. About two and half sunspot cycles are depicted here. The butterfly diagram shows that the sunspot emergence belt moves equatorward, whereas, the weak surface radial field moves poleward with the progress of the solar cycle. The polar field reversal starts at about the time of sunspot maximum, and the polar field is the strongest at sunspot minimum.}
\label{fig1}
\end{center}
\end{figure}

Sunspots have been observed for many years and systematically so from the early 17th Century since the invention of the telescope. The number of sunspots on the surface of the Sun varies cyclically going through successive maxima and minima with an average period of 11 years (Fig.~1-top). Given that sunspots are strongly magnetized regions (Hale 1908) with typical field strengths on the order of $1000$ Gauss (G), this 11 year solar cycle is essentially a magnetic cycle. At the beginning of the cycle, sunspots are observed to appear at high latitudes and with the progress of the cycle, more and more sunspots appear at lower and lower latitudes with the cycle ending close to the equator. This equatorward migration of the sunspot formation belt takes place in both the hemispheres generating the well known solar butterfly diagram of sunspots (Fig.~1-bottom). At the same time, the weaker, radial field outside of sunspots are seen to move poleward with the progress of the cycle -- reaching the poles and starting to reverse the old cycle polar field around the time of sunspot maxima. The polar field reaches its strongest intensity at sunspot minimum, thereby having a $90^{\circ}$ phase lag with the sunspot cycle. The polar field reversal occurs with an average period of 11 years as well and is therefore coupled to the sunspot cycle.

The solar cycle is not strictly periodic. There are variations from one cycle to another in both the period and more notably in the amplitude of the cycle. Even the length and nature of minima varies from one cycle to another, which was highlighted recently in the recent, unusually long solar cycle 23 minimum (see Supplementary Information--Fig.~1 from Nandy, Mu\~noz-Jaramillo \& Martens 2011). In the recorded history of sunspot observations there is a period between 1645--1715 AD when hardly any sunspots were seen on the solar surface (Fig.~1-top). This period is termed as the Maunder minimum and it is thought that the sunspot cycle stopped during this phase but restarted again. Solar activity reconstructions based on cosmogenic isotopes indicate that in the past, several such episodes have occurred in the Sun (Usoskin, Solanki \& Kovaltsov 2007). Such grand minima are therefore an integral part of solar activity over long timescales. Some evidence based on tree-ring data also suggests that there was a weak cycle -- at least in the solar open flux and polar fields -- during the Maunder minima (Miyahara et al.\ 2004).

The sunspots themselves are observed to emerge in pairs of opposite polarities (see Fig.~2). The bipolar sunspot pairs have a systematic tilt with pairs at higher latitudes being more tilted than those at low latitudes -- following what is known as the Joy's law distribution of tilt angles (Hale et al. 1919). It is also seen that the orientation of these bipolar sunspot pairs are always reversed across the equator, i.e., if the leading polarity spots on the Northern-hemisphere are positive, then the leading polarity spots in the Southern-hemisphere are negative. This relative orientation reverses from one $11$ year cycle to another and therefore the full magnetic cycle constitutes $22$ years.

\begin{figure}[t]
\begin{center}
\includegraphics[width=7.0cm]{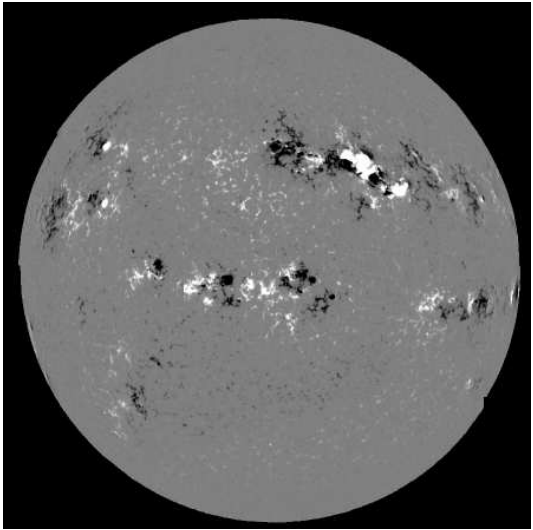}
\caption{A magnetogram showing the vertical component of the solar magnetic field on the Sun's surface. The typical strength of the magnetic field within sunspots is on the order of 1000 G. White regions denote sunspots which have positive vertical magnetic field and dark regions show sunspots with negative vertical magnetic field. Sunspots occur in bipolar pairs which have opposite polarities. Their polarity orientation is reversed across the equator, which is evident in the above figure. The line joining the two polarities of a sunspot pair are systematically tilted relative to the local parallel of latitude and this tilt increases with increasing latitude away from the equator.}
\label{fig2}
\end{center}
\end{figure}

It is to be noted that similar magnetic cycles -- exhibiting a wide range of activity behavior -- including Maunder minima like states are also observed in other solar-like stars (Baliunas \& Vaughan 1985; Giampapa et al.\ 2006; Poppenh\"ager et al.\ 2009). This indicates that magnetic cycles are a common and ubiquitous phenomena in stellar physics and plausibly have a common underlying mechanism at its origin (see e.g., Saar \ Brandenburg 1999, Nandy 2004).

Plasma flows in the Sun are believed to play a crucial role in the solar cycle and are now observed routinely. Tracking sunspots was an early indicator of the differential rotation of the Sun and it is known that the equator of the Sun rotates faster than its poles. Now, with the observational technique of helioseismology we can also observe the differential rotation in the solar convection zone (SCZ) and know that the mainly latitudinal differential rotation in the SCZ changes over to a radial differential rotation at its base and is concentrated in a thin layer known as the tachocline (Charbonneau et al. 1999). The differential rotation of the Sun constitutes the $\phi$-component of the plasma flow in the solar interior. There is also another component of the plasma flow in the Sun's interior in the meridional plane ($r$-$\theta$ plane); this meridional flow of plasma is observed to be poleward in the surface and near-surface layers (Gonz{\'a}lez Hern{\'a}ndez et al.\ 2006) and is believed to pervade the SCZ (Giles et al. 1997), with an equatorward counterflow theoretically expected at or below the base of the SCZ (Nandy \& Choudhuri 2002). Convective turbulence is also expected to be a feature of the plasma in the Sun's interior and visible manifestations of super-granular cells in the surface are an indicator of the underlying convective turbulent motions.

With these ideas of solar plasma flows and magnetic field dynamics in the background, I will trace the history of development of ideas in solar dynamo theory in the next section.

\section{Solar Dynamo Theory: Development of Ideas}

The behavior of magnetic fields in plasma systems (such as that in stellar interiors) is governed by the induction equation
\begin{equation}
\frac{\partial \bf{B}}{\partial t} =  {\nabla} \times (\bf{v} \times \bf{B} - \eta \, \nabla
\times \bf{B)},
\end{equation}
where {\bf{B}} is the magnetic field, {\bf{v}} the velocity field and $\eta$ the effective magnetic diffusivity of the system -- which in this case, encompasses the enhanced contribution due to turbulence in stelar interiors. Astrophysical plasma systems have a high characteristic magnetic Reynolds number (the ratio of the first to the second term on the R.H.S.\ of the above equation). The concept of flux-freezing holds in such systems (Alfv\'en 1942), wherein, the magnetic fields remain frozen in the fluid flow -- essentially because the diffusion timescale is much larger compared to the flow timescale. This allows the energy of convective flows in the solar convection zone (SCZ) to be drawn into producing and amplifying magnetic fields against dissipation -- which is the dynamo mechanism mechanism in a nutshell!

Assuming spherical symmetry (which is applicable to stellar interiors), the magnetic and velocity fields can be expressed as
\begin{equation}
{\bf B} = B_{\phi} {\bf \hat{e}}_{\phi} + \nabla \times (A {\bf \hat{e}}_{\phi})
\end{equation}
\begin{equation}
{\bf v} = r\sin(\theta)\Omega{\bf \hat{e}}_{\phi} + {\bf v}_p.
\end{equation}
The first term on the R.H.S.\ of Eqn.~$2.2$ is the toroidal component (i.e, in the $\phi$-direction) and the second term is the poloidal component (i.e., in the $r$-$\theta$ plane) of the magnetic field. In the case of the velocity field (Eqn.~2.3), these two terms are the rotation $\Omega$ and meridional circulation $v_p$, respectively. Due to the differentially rotating Sun, any pre-existing poloidal field would get stretched in the direction of rotation creating a toroidal component. It is believed that these strong toroidal flux tubes are stored and amplified in the tachocline region -- which coincides with a region of sub-adiabatic temperature gradient known as the overshoot layer and where magnetic buoyancy is suppressed (Spiegel \& Weiss 1980; van Ballegooijen 1982). These horizontal toroidal flux tubes in the solar interior become unstable to magnetic buoyancy (Parker 1955a) when they come out in to the SCZ (through overshooting turbulence or meridional up-flows) and erupt through the solar surface creating bipolar sunspot pairs. During their buoyant rise, these flux tubes acquire tilt due to the Coriolis force generating the Joy's law distribution of tilt angles (D'Silva \& Choudhuri 1993).

For the dynamo chain of events to be to be completed, the toroidal field has to regenerate the poloidal component so that the solar cycle can go on through a recycling of magnetic flux between the two components (in Eqn.~2.2) mediated via the plasma flows. The first mechanism proposed as a regeneration mechanism for the poloidal component of the magnetic field is the action of small-scale helical turbulence on rising magnetic flux tubes (Parker 1955b) -- which twists the toroidal field back into the meridional place (thereby generating a $r$-$\theta$ or poloidal component). This mechanism relying on helical turbulence is traditionally known as the mean-field $\alpha$-effect.

Simulations of the buoyant rise of toroidal flux tubes showed that the field strength of these flux tubes have to be close to $10^5$ G at the base of the SCZ to match the emergence properties of bipolar sunspot pairs (Choudhuri \& Gilman 1987) and their tilt angle distribution (D'Silva \& Choudhuri 1993; Fan, Fisher \& DeLuca 1993). The required field strength is much stronger than the equipartition magnetic field strength in the SCZ (at which the energy in the fields and convection are in equipartition). This called into question the viability of the mean-field $\alpha$-effect which is likely to get quenched by such strong fields.

\begin{figure}[t]
\begin{center}
\includegraphics[width=10.0cm]{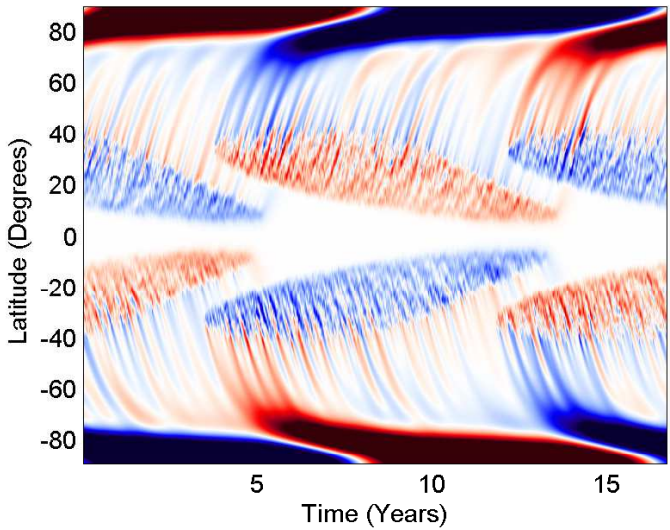}
\caption{A simulated solar butterfly diagram showing the evolution of the Sun's radial magnetic field at the surface from a recently developed kinematic solar dynamo model. Eruptions of bipolar sunspot pairs are evident at lower latitudes; the redistribution of their flux through surface flux transport processes involving meridional circulation, differential rotation and diffusion generates a large-scale global dipolar field. This cancels the older cycle polar fields at high latitudes and creates the polar fields of the new cycle. The observational features of the solar cycle are reproduced well through such simulations. This particular simulation is from a kinematic dynamo model that includes a realistic algorithm for the emergence of tilted bipolar sunspot pairs (Mu\~noz-Jaramillo, Nandy \& Martens 2010).}
\label{fig3}
\end{center}
\end{figure}

An alternative mechanism, originally proposed by Babcock (1961) and Leighton (1969), is now the leading candidate for the regeneration of the poloidal component of the magnetic field. In this mechanism, the decay of tilted bipolar sunspot pairs, mediated via diffusion and meridional circulation creates a large-scale solar dipole from the emergence of individual bi-poles with a non-zero tilt (which have a net dipole moment). Thus the large-scale solar poloidal field is created and manifests as the global solar dipole configuration -- most intense at solar minimum -- from which the next cycle toroidal field is generated. This keeps the cycle going. This observed process of surface flux distribution is reproduced by surface flux transport simulations (Mackay \& Lockwood 2002; Jiang et al. 2011) and dynamo models (Mu\~noz-Jaramillo, Nandy \& Martens 2010) based on the Babcock and Leighton (BL) idea. More importantly, recently inferred correlations between observed proxies of the BL source term and the strength of the next sunspot cycle (Dasi-Espuig et al. 2010) lend strong support to the BL mechanism for poloidal field creation.

While full magnetohydrodynamic (MHD) simulations of the solar interior are becoming more and more sophisticated due to rapid improvements in our numerical capabilities they are still not capable of reproducing the rich variety of solar cycle observations. Kinematic dynamo models play a useful role here, wherein, their simplicity allows for a transparent underlying physics, and the capability to explore the full breadth of solar cycle properties through numerical investigations. In these models, the plasma flows are prescribed (and is not a dynamical variable) and the magnetic induction equation (Eqn.~2.1) is solved with suitable boundary conditions and parameterizations and algorithms necessary for accounting for the physics of the solar interior. Kinematic flux transport dynamo models (as opposed to models based on dynamo waves) rely on the plasma flows to transport flux between different source layers for the field components and have now become the leading tool for exploring solar activity (Jouve et al.\ 2008; see also the reviews by Charbonneau 2010) and recently developed models are good enough to reproduce not only the global features of the solar cycle, but also the dynamics of surface flux transport and polar field reversal accurately (Fig.~3).

We do not provide a pedagogical description of the mean field dynamo equations here and instead refer (interested) readers to Nandy (2010a) -- where a more complete description is provided, and to Nandy (2010b) -- where some issues of current interest are discussed that will complement this review. I now jump ahead and devote the next section to discussing what I deem to be fertile grounds for dynamo theory in the coming years.

\section{A Future Outlook}

It is interesting times for solar dynamo theory. In the last decade major improvements have happened in the field of kinematic dynamo modeling and the so-called ``flux transport'' (kinematic) dynamo models have become useful tools for explaining various features of the solar cycle. {\it{I must note here in passing that I do not necessarily endorse the usage of the term ``flux-transport'' in this context as many dynamo modelers who declare their models to be ``flux-transport'' dynamo models do not include either magnetic buoyancy or downward pumping of magnetic flux -- which are plausibly the most efficient of all flux transport mechanisms in the solar interior}}. Simultaneously, in the last several years or so, full MHD models have developed to the point where structured large-scale magnetic fields (Brown et al. 2011) and cyclic reversals in solar-like conditions are starting to be seen (Ghizaru, Charbonneau \& Smolarkiewicz 2010). With the available long-term sunspot record in conjunction with polar magnetic field measurements stretching over three cycles and  helioseismic observations covering at least a full cycle, we also now have the requisite observational data to constrain and indeed, drive dynamo simulations. I have no doubt that the most important advances in dynamo theory are going to come at the confluence of all of these, where kinematic and full MHD simulations of the solar cycle complement each other and assimilate magnetic field and plasma flow data. Below, I elaborate on some specific aspects -- which are important and which I believe are going to garner sustained attention in the near future.

\subsection{It takes Two to Tango: Kinematic Dynamos and full MHD Simulations}

The history of solar dynamo theory has been such that kinematic dynamo modeling and full MHD simulations of solar and stellar interiors have often progressed in parallel with very limited feedback from each other. This is changing. An prominent example is explorations of the nature of the meridional flow which is largely unconstrained by observations. Kinematic dynamo models, especially those based on the BL mechanism -- wherein the source layers for the poloidal and toroidal field are spatially separated, find that the meridional circulation is an important factor in determining various aspects of the solar cycle, including fluctuations in the period (Charbonneau \& Dikpati 2000) and the latitudinal distribution of sunspots (Nandy \& Choudhuri 2002); the latter proposed that the flow has to penetrate below the base of the SCZ for reproducing the observed solar butterfly diagram. While this suggestion was controversial at the time, full MHD models are not beginning to find predominantly unicellular flows (averaged over solar cycle timescales) such as that used in kinematic dynamo models with overshooting plumes and plasma flows that can indeed penetrate below the base of the SCZ (Rogers, Glatzmaier \& Jones 2006; Garaud \& Brummell 2008). This penetration of plasma plumes and flows into the radiative interior is important because this region is stable to magnetic buoyancy; this in conjunction with the longer diffusion timescales there, allow for magnetic field storage and slow transport (by the equatorward meridional counterflow) that are relevant for establishing an 11 year solar cycle. Clearly, this convergence of kinematic dynamo models with MHD simulations of the solar interior reiterate the importance of meridional circulation vis-a-vis the solar cycle.

It is to be noted that kinematic dynamo models does not necessarily self-consistently include the full range of physics possible in stellar interiors and therefore must assimilate those physical processes that full MHD simulations point out to be important. A case in point is the downward pumping of magnetic flux in the presence of rotating, stratified convection -- a process that is very efficient at pumping magnetic fields down into the stable interior below the base of the SCZ (Tobias et al. 2001; K\"apyl\"a et al.\ 2006). Only few kinematic dynamo models include downward pumping -- those that do show that it has a significant influence on the dynamo (Guerrero \& de Gouveia Dal Pino 2008); evidently, given its low timescales, the process of flux pumping competes with meridional flow and turbulent diffusion in transporting flux between the source layers of the dynamo and should be considered in kinematic dynamo models. I believe useful exchanges between kinematic dynamo modeling and full MHD simulations of the kind highlighted above are going to lead to more meaningful advances in the understanding of the solar cycle.

\subsection{Beyond Normal: Irregularities in the Solar Cycle}

Fig.~1 shows that the amplitude of the solar cycle varies significantly. While magnetic buoyancy presumably sets the order of magnitude field strength (Nandy 2002), smaller fluctuations around these values likely have a different origin. Then there are grand minima phases like the Maunder minimum when the sunspot cycle basically stops. There are also variations in the period of the cycle, the length of the minimum between solar cycles and the strength of the polar field. Given that dynamo models are now reasonably successful in reproducing and explaining the regular features of the solar cycle, the challenge ahead lies in confronting models of magnetic field generation with irregularities in the solar cycle.

What causes these variations -- is it due to stochastic fluctuations, e.g., in the meridional flow and the dynamo $\alpha$-effect (Charbonneau \& Dikpati 2000), or is it the non-linear feedback of the fields on the flows (Tobias 1997) that may lead to chaotic modulation or is it due to sudden changes in some dynamo ingredient like the meridional flow (Karak 2010). Some observational analysis seems to rule out the existence of low-dimensional chaos in the solar cycle (Mininni, G\'omez \& Mindlin 2002), however, the underlying non-linearities of the system can certainly exhibit chaotic modulation when the forcing is strong; the jury is still out on this one.

While it is tempting to surmise that the same process(es) that produce cycle to cycle fluctuations also sometimes bring down the activity levels to produce grand minima like episodes -- that would be intellectually satisfying -- does it hold in reality? The unusually long minimum of solar cycle 23 and a dynamo based explanation of this episode (Nandy, Mu\~noz-Jaramillo, \& Martens 2011) suggests that such variations in the nature of solar minima are caused by different processes than those that cause grand minima episodes.  The way the Sun gets into Maunder-minima like episodes and the way it gets out from these quiescent phases also holds keys to understanding many of the subtleties of the dynamo process. We are currently just scratching the surface here and need to know more about the origin and implications of solar cycle fluctuations and extreme events.

\subsection{Data Assimilation and Solar Cycle Forecasting}

The holy grail of solar dynamo theory is to be able to forecast the solar cycle. It is easier said than done, especially given the uncertainties regarding the relative roles of flux transport processes such as diffusion, circulation and downward flux pumping in the solar interior. Dynamo based predictions for the next cycle do not converge (Dikpati, de Toma \& Gilman 2006; Choudhuri, Chatterjee \& Jiang 2007) and this is due to differing importance laid on meridional circulation and turbulent diffusion (see Yeates, Nandy \& Mackay 2008). The memory of the solar cycle is related to the timescale of the dominant flux transport process and since the build up of memory makes prediction possible, accurate predictions will require understanding the subtleties of various flux transport processes in the solar interior. Prediction also requires the  assimilation of data. At the least, the data could be polar field (or surface magnetic field) observations -- which are precursors of the next cycle (Dasi-Espuig et al.\ 2010) -- and which can be fed into kinematic dynamo models to generate the toroidal field of the next cycle.

However, more sophisticated treatment of the prediction problem, including predictions of solar cycle timing would also necessitate the realistic assimilation of observed plasma flow data in solar dynamo models; such efforts are just beginning (see Mu\~noz-Jaramillo, Nandy \& Martens 2009). Efforts are also on to take methods used in meteorology and develop them for solar cycle predictions (see e.g., Katiashvili \& Kosovichev 2008 and Jouve, Brun \& Talagrand 2011). Given the success of such techniques in weather prediction, this would seem to a pragmatic approach. In this approach, an observed state of the Sun is integrated forward in time using a solar model, and continuous corrections based on real-time observations are made to keep the model on track. Evidently, such an approach requires a good dynamo model to begin with and techniques for assimilating both plasma flow and magnetic field observations within this model. Given the need of reliable space weather forecasts, it would seem that attempts to predict the solar cycle would also drive the motivation to understand it better in the coming years!

\subsection{Solar-Stellar Connections: Stellar Activity and Dynamo Theory}

The Sun is one amongst many stars with similar properties. There is no reason than that a dynamo model for magnetic field generation would be unique to the Sun; in fact, one would think that the Sun just offers one data point for the dynamo, whilst observations of other solar-like stars offer multiple data points -- in the parameter space of e.g., differential rotation or convective turn-over time -- to constrain the dynamo mechanism with. An extreme form of this argument would imply that an acceptable dynamo model for the Sun should also be able to explain the activity of other stars with suitably adjusted parameters conforming to the interior of these stars. Typically however, not all dynamo models are confronted with stellar observational data. This is because the stellar data does not have well-constrained plasma flow observations (differential rotation and meridional circulation) and in the absence of these, there are too many free parameters in spatially extended dynamo models for the whole exercise to be meaningful. Often, low-order mathematical models have been used instead to explore the wide range of stellar activity observations and they provide useful insight on how the exhibited dynamics is related to the underlying structure of the dynamo equations (see e.g., Wilmot-Smith et al.\ 2005, 2006 and references therein).

Instead of modeling individual, unresolved stars, an alternative and promising approach is to extend solar dynamo models to explore well-known relationships that exist in the stellar data, e.g., the rotation rate, activity amplitude and period relationship (Noyes, Vaughan \& Weiss 1984; Brandenburg, Saar \& Turpin 1998; Nandy 2004; Jouve, Brown \& Brun 2010). Surface flux transport ideas gleaned from the Sun can also be adapted to explain the formation of polar caps and starspot dynamics (Schrijver \& Title 2001; Holzwarth, Mackay \& Jardine 2006; Isik, Sch\"ussler \& Solanki 2007). Such approaches sometimes throw up surprises; it appears that the BL dynamo as envisaged for the Sun runs into some difficulty when confronted with the activity-period relationship from stellar observations (Jouve, Brown \& Brun 2010). Given that the BL dynamo is supported both by theoretical models and solar observations are we missing something when we confront such models with stellar data -- perhaps a realistic handling of downward flux pumping? A combination of full MHD simulations of stellar interiors and kinematic dynamo modeling would be required to address such vexing issues; certainly more effort should be devoted to using stellar activity relationships to constrain dynamo theory more meaningfully.

\subsection{Space Climate: The Solar Dynamo in Time}

I would like to end by bringing up a topic that has not received much attention but which is going to become important in the years to come. This pertains to uncovering the long-term evolution of the solar dynamo over the life-time of the Sun and using this in conjunction with ideas of solar luminosity variations with radius and age to come up with a complete profile of the radiative history of the Sun. Theoretical and observational studies of solar-like stars at different main-sequence ages are going to play an important role in this regard (Nandy \& Martens 2007). With increasing importance of understanding the causes and consequences of global climate change, the need to figure out how our parent star has shaped our atmosphere and climate over time will also become an important quest.\\

{\it{I would like to acknowledge the Department of Science and Technology of the Government of India for supporting my research through the Ramanujan Fellowship. I am grateful to the Argentinians -- specifically for being wonderful hosts, and in general -- for giving tango to the world.}}

\label{lastpage}


\begin{thebibliography}{}

\bibitem[{{}()}]{}
Alfv\'en, H. 1942, {\it{Ark.\ f.\ Mat.\ Astr.\ o Fysik}}, 29B, No.\ 2

\bibitem[{{}()}]{}
{Babcock}, H.~W. 1961, \textit{ApJ}, 133, 572

\bibitem[{{}()}]{}
Baliunas, S.L., \& Vaughan, A.H. 1985, \textit{ARAA}, 23, 379

\bibitem[{{}()}]{}
Brandenburg, A., Saar, S.H., \& Turpin, C.R. 1998, \textit{ApJ}, 498, L51

\bibitem[{{}()}]{}
Brown, B.P. et al.\ 2011, \textit{ApJ}, 731, 69

\bibitem[{{}()}]{}
{Charbonneau}, P. 2010, {\it{Living Reviews in Solar Physics}}, 7, 3

\bibitem[{{}()}]{}
Charbonneau, P. et al.\ 1999, \textit{ApJ}, 527, 445

\bibitem[{{}()}]{}
Charbonneau, P. \& Dikpati, M. 2000, \textit{ApJ}, 543, 1027

\bibitem[{{}()}]{}
Choudhuri, A.R., \& Gilman, P.A. 1987, \textit{ApJ}, 316, 788

\bibitem[{{}()}]{}
Choudhuri, A. R., Chatterjee, P., \& Jiang, J. 2007, \textit{Phys. Rev. Lett.}, 98, 131103

\bibitem[{{}()}]{}
Dasi-Espuig, M. et al.\ 2010, \textit{A\&A}, 518, A7

\bibitem[{{}()}]{}
Dikpati, M., de Toma, G., \& Gilman, P. A. 2006, {\it{Geophys.\ Res.\ Lett.}}, 33, 5102

\bibitem[{{}()}]{}
{D'Silva}, S. \& {Choudhuri}, A.~R. 1993, \textit{A\&A}, 272, 621

\bibitem[{{}()}]{}
{Fan}, Y., {Fisher}, G.~H. \& {Deluca}, E.~E. 1993, \textit{ApJ}, 405, 390

\bibitem[{{}()}]{}
Garaud, P., \& Brummell, N.H. 2008, \textit{ApJ},674, 498

\bibitem[{{}()}]{}
Ghizaru, M., Charbonneau, P., \& Smolarkiewicz, P.K. 2010, \textit{ApJ}, 715, L133

\bibitem[{{}()}]{}
Giampapa, M.S., Hall, J.C., Radick, R.R., Baliunas, S.L. 2006, \textit{ApJ}, 651, 444

\bibitem[{{}()}]{}
Giles et al. 1997, \textit{Nature}, 390, 52

\bibitem[{{}()}]{}
{Gonz{\'a}lez Hern{\'a}ndez}, I., et al.\ 2006, \textit{ApJ}, 638, 576

\bibitem[{{}()}]{}
Guerrero, G., \& de Gouveia Dal Pino, E.M. 2008, \textit{A\&A}, 484, 267

\bibitem[{{}()}]{}
Hale, G.E. 1908, \textit{ApJ}, 28, 315

\bibitem[{{}()}]{}
Hale, G.E., Ellerman, F., Nicholson, S.B., \& Joy, A.H. 1919, \textit{ApJ}, 49, 153

\bibitem[{{}()}]{}
Holzwarth, V., Mackay, D.H., \& Jardine, M. 2006, \textit{MNRAS}, 369, 1703

\bibitem[{{}()}]{}
Isik, E., Sch\"ussler, M., Solanki, S.K. 2007, \textit{A\&A}, 464, 1049

\bibitem[{{}()}]{}
Jiang, J., Cameron, R.H., Schmitt, D., \& Sch\"ussler, M. 2011, \textit{A\&A}, 528, A83

\bibitem[{{}()}]{}
{Jouve}, L., et al.\ 2008, \textit{A\&A}, 483, 949

\bibitem[{{}()}]{}
Jouve, L., Brown, B.P., \& Brun, A.S. 2010. \textit{A\&A}, 509, A32

\bibitem[{{}()}]{}
Jouve, L., Brun, A.S., \& Talagrand, O. 2011, \textit{ApJ}, 735, 31

\bibitem[{{}()}]{}
K\"apyl\"a, P.J., Korpi, M.J., Ossendrijver, M., \& Stix, M. 2006, \textit{A\&A}, 455, 401

\bibitem[{{}()}]{}
Karak, B.B. 2010, ApJ, 724, 1021

\bibitem[{{}()}]{}
Kitiashvili, I., \& Kosovichev, A.G. 2008, \textit{ApJ}, 688, L49

\bibitem[{{}()}]{}
{Leighton}, R.~B. 1969, ApJ, 156, 1

\bibitem[{{}()}]{}
Mackay, D.H., \&, Lockwood, M. 2002, \textit{Solar Phys.}, 209, 287

\bibitem[{{}()}]{}
Mininni, P.D., G\'omez, D.O., \& Mindlin, G.B. 2002, \textit{Phys. Rev. Lett.}, 89, 061101

\bibitem[{{}()}]{}
Miyahara, H. et al.\ 2004, \textit{Solar Phys.}, 224, 317

\bibitem[{{}()}]{}
Mu\~noz-Jaramillo, A., Nandy, D., \& Martens, P. C. H. 2009, ApJ, 698, 461

\bibitem[{{}()}]{}
Mu\~noz-Jaramillo, A., Nandy, D., Martens, P.C.H. \& Yeates, A.R. 2010, ApJL, 720, L20

\bibitem[{{}()}]{}
{Nandy}, D. 2002, Astrophysics and Space Science, 282, 209

\bibitem[{{}()}]{}
Nandy, D. 2004, \textit{Solar Phys.}, 224, 161

\bibitem[{{}()}]{}
Nandy, D. 2010a, in Heliophysical Processes, ed. N.~Gopalswamy, S.~Hasan \& A.~Ambastha, Springer (Berlin), 35

\bibitem[{{}()}]{}
Nandy, D. 2010b, in Magnetic Coupling between the Interior and Atmosphere of the Sun, ed. S. S. Hasan \& R. J. Rutten, Springer (Berlin), 86

\bibitem[{{}()}]{}
{Nandy}, D. \& {Choudhuri}, A.~R. 2002, Science, 296, 1671

\bibitem[{{}()}]{}
Nandy, D., \& Martens, P.C.H. 2007, \textit{Adv. Sp. Res.}, 40, 891

\bibitem[{{}()}]{}
Nandy, D., Mu\~noz-Jaramillo, A. \& Martens, P. C. H. 2011, \textit{Nature}, 471, 80

\bibitem[{{}()}]{}
Noyes, R.W., Weiss, N.O., \& Vaughan, A.H. 1984, \textit{ApJ}. 287, 769

\bibitem[{{}()}]{}
{Parker}, E.~N. 1955a, ApJ, 121, 491

\bibitem[{{}()}]{}
{Parker}, E.~N. 1955b, ApJ, 122, 293

\bibitem[{{}()}]{}
Poppenh\"ager, K., Robrade, J., Schmitt, J.H.M.M., \& Hall, J.C. 2009, \textit{A\&A}, 508, 1417

\bibitem[{{}()}]{}
Rogers, T.M., Glatzmaier, G.A., \& Jones, C.A. 2006, \textit{ApJ}, 653, 765

\bibitem[{{}()}]{}
Saar, S.H., \& Brandenburg, A. 1999, \textit{ApJ}, 524, 295

\bibitem[{{}()}]{}
Schrijver, C.J., \& Title, A.M. 2001, \textit{ApJ}, 551, 1099

\bibitem[{{}()}]{}
Spiegel, E.A. \& Weiss, N.O. 1980, \textit{Nature}, 287, 616

\bibitem[{{}()}]{}
Tobias S.M. 1997, \textit{A\&A}, 322, 1007

\bibitem[{{}()}]{}
{Tobias}, S.~M., {Brunnell}, N.H., {Clune}, T.L. \& {Toomre}, J. 2001, ApJ, 549, 1183

\bibitem[{{}()}]{}
Usoskin, I.G., Solanki S.K., \& Kovaltsov, G.A. 2007, \textit{A\&A}, 471, 301

\bibitem[{{}()}]{}
van Ballegooijen, A.~A. 1982, \textit{A\&A}, 113, 99

\bibitem[{{}()}]{}
Wilmot-Smith, A.~L., et al. 2005, MNRAS, 363, 1167

\bibitem[{{}()}]{}
Wilmot-Smith, A.~L., et al. 2006, ApJ, 652, 696

\bibitem[{{}()}]{}
{Yeates}, A.~R., {Nandy}, D., \& {Mackay}, D.~H. 2008, ApJ, 673, 544

\end{thebibliography}
\end{document}